\documentclass[reprint,amsmath,amssymb,aps,prl,superscriptaddress, showpacs, eprint]{revtex4-1}
\usepackage[utf8]{inputenc}

\usepackage{amsmath,amssymb,ascmac,fancybox}
\usepackage{color}
\usepackage{graphicx}
\usepackage{url}
\usepackage{siunitx}
\usepackage{bm}
\usepackage{physics}
\usepackage{mathtools}
\usepackage{hyperref}
\usepackage[T1]{fontenc}
\usepackage{ulem}
\graphicspath{figures/}

\renewcommand{\vec}{\vb*}

\newcommand{\vol}{\Omega}

\newcommand{\eff}{\vartheta}
\newcommand{\HPower}{\eta}

\newcommand{\dc}{\mathrm{dc}}
\newcommand{\ac}{\mathrm{ac}}

\newcommand{\Pin}{P_{\mathrm{in}}}

\begin{document}
\title{High-efficiency energy harvesting based on nonlinear Hall rectifier 
}
\date{\today}
\author{Yugo Onishi}
\affiliation{Department of Physics, Massachusetts Institute of Technology, Cambridge, MA 02139, USA}
\author{Liang Fu}
\affiliation{Department of Physics, Massachusetts Institute of Technology, Cambridge, MA 02139, USA}

\begin{abstract}
    Noncentrosymmetric quantum materials can convert AC input current into DC transverse current through the nonlinear Hall effect at zero magnetic field. We analyze the AC-DC power conversion efficiency of such ``Hall rectifier'' and suggest its application in wireless charging and energy harvesting. Our key observation is that the development of Hall voltage results in a change of longitudinal resistance, resulting in a violation of Ohm's law due to the nonlinear Hall effect. This feedback mechanism balances the input power and the output power and hence is crucial to understanding the power transfer from source to load. We derive a general expression for the power conversion efficiency in terms of material parameters, external load resistance, and input power. As the Hall current is perpendicular to the electric field and does not generate Joule heating by itself, %
    we obtain high power conversion efficiency %
    when the Hall angle (which increases with the input power) is large and the load resistance is optimized. %
    Promising materials for high-efficiency Hall rectifiers are also discussed. 
    
\end{abstract}

\maketitle

Technological applications of electromagnetic waves are essential to modern society. The development and commercialization of 5G wireless network has created a lot of interest in exploring emerging applications in millimeter wave (mmWave) and terahertz  spectrum~\cite{tripathi_millimeter-wave_2021, tonouchi_cutting-edge_2007}. 
Owing to the shorter wavelength of millimeter waves, 
compact-size antennas can receive power with a higher efficiency, promising for wireless powered Internet of Things (IoT) devices ~\cite{tripathi_millimeter-wave_2021, akyildiz_internet_2010}. Conversion of THz waves into usable power is important for energy harvesting, since the solar energy reaching the earth is re-radiated in the form of infrared light (20-40 THz) which is wasted.  %

A key ingredient for mm-Wave and THz technology is the conversion of electromagnetic waves into direct current (DC) electricity. This can be achieved with a diode combined with an antenna, called rectenna.  
Using its nonlinear $I-V$ characteristic, the diode rectifies the alternating current induced in the antenna by electromagnetic fields into direct current, to produce DC power. %
Rectenna have been extensively investigated for wireless power transmission via microwave radiation  ~\cite{donchev_rectenna_2014, moddel2013}.
While  the power conversion efficiency reaches as high as $\SI{80}{\percent}$ at frequencies $\lesssim\SI{10}{GHz}$~\cite{donchev_rectenna_2014}, the realization of mmWave and terahertz rectenna is highly challenging. A main difficulty is to design diodes operating efficiently at such high frequencies, as conventional diodes have a cutoff frequency determined by RC time constant of semiconductor junctions.
The RF-DC conversion efficiency of a rectenna with GaAs Schottky diode is only around $\SI{2}{\percent}$ at $300$GHz ~\cite{mizojiri_recent_2019}. 
Also, conventional diodes have a minimum threshold voltage and therefore cannot work efficiently at low input power for ambient energy harvesting~\cite{lu_wireless_2015, okba_compact_2019, hemour_towards_2014}. This bottleneck hinders, for example,  the development of efficient wake-up circuits in IoT, which should work with no bias voltage. %

Recently, a new physical principle for current rectification has been proposed based on the {\it intrinsic, nonreciprocal} response of noncentrosymmetric quantum materials at zero magnetic field~\cite{isobe_high-frequency_2020}. 
In particular, a  {\it transverse} current can be induced in a noncentrosymmetric conductor by an applied electric field to the second order, known as the nonlinear Hall effect (NLHE)~\cite{Sodemann2015, ma_observation_2019, kang_nonlinear_2019, gao_field_2014}. When an AC electric field is applied at frequency $\omega$, NLHE simultaneously gives rise to an AC current at frequency $2\omega$ and a DC current. NLHE enables the generation of DC electricity from AC input by utilizing quantum electronic properties of noncentrosymmetric materials, without invoking p-n or metal-semiconductor junction. This quantum rectification process can also be viewed as a special type of linear photogalvanic effect due to Drude absorption rather than interband transition~\cite{deyo_semiclassical_2009, moore_confinement-induced_2010, matsyshyn_nonlinear_2019}. 

While the NLHE-based rectification has the potential to overcome the fundamental limitations of the conventional diode at high operation frequency and low input power, its AC-to-DC power conversion efficiency---one of the most important figures of merit---remains to be understood. 
In order to develop the NLHE-based energy harvesting and wake-up circuit technologies, it is essential to determine the power conversion efficiency and find the design principle for nonlinear Hall rectifier.

\begin{figure}
    \centering
    \includegraphics[width=0.9\columnwidth]{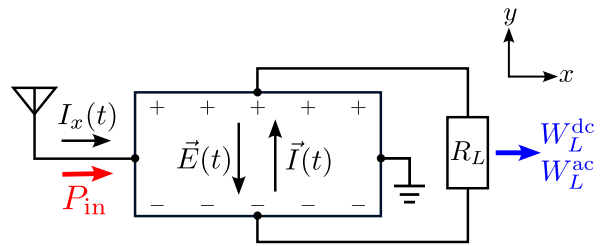}
    \caption{Schematic figure of rectenna based on nonlinear Hall effect. Antenna receives electromagnetic fields converting them into an AC electric current. The produced AC electric current flows into the conductor, where the AC current $I_x(t)$ is rectified into a DC current $I_y$ via NLHE along with an AC current with frequency $2\omega$. The induced DC current and AC current in $y$ direction flow into the external load $R_L$ and the power $W_L^\dc$ and $W_L^\ac$ are delivered.}
    \label{fig:schematics_rectenna}
\end{figure}

In this work, we study the power conversion efficiency of noncentrosymmetric conductors that convert AC input power into DC power through NLHE. 
To this end, we consider a circuit that consists of an antenna, a noncentrosymmetric conductor and an external load. The electromagnetic radiation received by the antenna induces an AC current, which flows into a noncentrosymmetric conductor.  The conductor (``Hall rectifier'')  induces a DC transverse current through NLHE, and the current flows into an external load to do work, as shown in Fig.~\ref{fig:schematics_rectenna}. %

Importantly, we show that in the power conversion through NLHE, the development of Hall field necessarily results in a change of longitudinal conductivity in the Hall rectifier, which is manifested as a violation of Ohm's law in the longitudinal $I-V$ relation due to NLHE. The feedback from the Hall field to the longitudinal response balances the input power and the output power, and hence is crucial in understanding the energy flow from the power source to the external load. Furtheremore, we derive the general expression for the power conversion efficiency in terms of material parameters, external load resistance, and input power. Remarkably, we show that the maximum power conversion efficiency reaches nearly \SI{100}{\percent} when the nonlinear Hall angle is sufficiently large and the load resistance is optimized in proportion to the internal resistance, which is a consequence of the nondissipative nature of the Hall current.

A key feature of NLHE is that the Hall current is always transverse to the electric field $\vec{E}$, and therefore does not by itself produce Joule heating.
When an electric field $\vec{E}$ is applied to a noncentrosymmetric conductor possessing Berry curvature in $\vec{k}$ space, the anomalous velocity leads to a Hall current $\vec{j}^{H}$ transverse to $\vec{E}$:
\begin{align}
{\vec{j}}^H &= \frac{e^2}{\hbar} \sum_{\vec{k}}   \left( \vec{E} \times \vec{\Omega}_{\vec{E}}(\vec{k})  \right) f_{\vec{E}}({\vec{k}}).  \label{eq:jH}
\end{align}
Here $f_{\vec{E}}(\vec{k})$ is the distribution function in the current carrying state driven by the electric field, and $\vec{\Omega}_{\vec{E}}(\vec{k})$ is the Berry curvature in the presence of the electric field $\vec{E}$. The anomalous velocity due to Berry curvature results in NLHE in several ways. In the presence of time reversal symmetry, the leading contribution to the Hall current ${\vec{j}}^H$ is second order in $\vec{E}$, as it comes from the product of electric field induced change in distribution function and the anomalous velocity, both of which are at first order in $\vec{E}$. This contribution can be described with the Berry curvature dipole~\cite{Sodemann2015}. Another contribution can arise from the change of Berry curvature due to external electric field. The change of $\vec{\Omega}_{\vec{E}}(\vec{k})$ at first order in  $\vec{E}$ contributes to second-order NLHE through Eq.~\eqref{eq:jH} with unperturbed distribution function. This contribution is present in time-reversal and inversion breaking systems and comes from the quantum metric~\cite{gao_field_2014}. 
Since its theoretical prediction~\cite{Sodemann2015,gao_field_2014}, the NLHE has been intensively studied and experimentally observed in various quantum materials~\cite{du_nonlinear_2021, ma_observation_2019, he_giant_2021, kang_nonlinear_2019, shvetsov_nonlinear_2019, gao2022quantummetricNLHE, zhang_terahertz_2021, kumar_room-temperature_2021, shi_symmetry_2019, ma_anomalous_2022, tiwari_giant_2021, zhang_giant_2022-1, zhang_electrically_2018, zhang_giant_2022, Tokura2018, michishita_dissipation_2022, ideue_symmetry_2021, kang_observation_2018, kang_switchable_2022, wang_intrinsic_2021, xiao_electrical_2020, toshio_plasmonic_2022, toshio_anomalous_2020, qin_strain_2021}. 
It %
is the dominant mechanism for second-order nonreciprocal transport at the leading order in the relaxation time $\tau$. %

To make the discussion concrete, let us consider a noncentrosymmetric crystal which is invariant under reflection $x\rightarrow -x$ and $z\rightarrow -z$, and has a polar axis along $y$ direction. Based on symmetry consideration, the current density ${\vec{j}}=(j_x, j_y)$ in this polar crystal induced by an electric field ${\bm E}=(E_x, E_y)$ can be expanded in powers of $\bm E$:
\begin{eqnarray}
j_x &=& \sigma_{xx} E_x + \sigma_{xyx} E_x E_y + \sigma_{xxxx} E_x^3 + ... , \label{eq:jx} \\
j_y &=& \sigma_{yy} E_y + \sigma_{yxx} E_x^2+ ...  \label{eq:jy}
\end{eqnarray}
Here $\sigma_{xx}$ and $\sigma_{yy}$ are the linear conductivities in $x$ and $y$ direction respectively. $\sigma_{yxx}$ and $\sigma_{xyx}$ are symmetry-allowed second-order conductivities. $\sigma_{xxxx}$ is a third-order conductivity which describes the change of conductivity proportional to $E_x^2$.  
Throughout this work, we consider the case that the externally applied electric field $E_x$ is along $x$ direction. Then, Eq.~\eqref{eq:jx} and \eqref{eq:jy} are exact up to the third order in $E_x$ as we will show below. All other symmetry allowed terms, such as the $\sigma_{yyy}E_y^2$ term in $j_y$~\cite{isobe_high-frequency_2020}, are at least fourth order in $E_x$. We consider the case that $E_x$ is not too large, so that additional terms not shown explicitly in Eq~\eqref{eq:jx} and \eqref{eq:jy} can be neglected.

Note that the second-order current from the nonlinear Hall effect is always transverse to the electric field regardless of its direction, as shown in Eq~\eqref{eq:jH}. It then follows from $\vec{j}^H \cdot \vec{E}= \sigma_{xyx}E_x^2 E_y + \sigma_{yxx} E_x^2 E_y=0$ that  
\begin{align}
\sigma_{yxx} = -\sigma_{xyx}. \label{eq:asymmetric_tensor}
\end{align}
Importantly, we note that $\sigma_{xyx}$ also represents the change of conductivity along the $x$ direction by the transverse electric field $E_y$, which may be called the ``bulk electric field effect''. 
It is thus remarkable that NLHE also provides a direct mechanism for the bulk electric field effect. 

When $\sigma_{yxx} = -\sigma_{xyx}$, the power dissipated in the nonlinear Hall conductor $Q$ is given by
\begin{eqnarray}
Q/\Omega= {\bm j} \cdot {\bm E} =\sigma_{xx} E_x^2 +\sigma_{yy} E_y^2 + \sigma_{xxxx} E_x^4, \label{eq:dissipated_power}
\end{eqnarray}
where $\vol=L_xL_yL_z$ is the volume of the device. $Q$ must be positive for any $\bm E$. When $\sigma_{xxxx} \geq 0$, the requirement is automatically satisfied. When $\sigma_{xxxx}<0$ or $\sigma_{xyx}\neq -\sigma_{yxx}$,  $Q>0$ is satisfied when the electric field is not too large, the regime considered in this work. 

\begin{figure}
    \centering
    \includegraphics[width=0.9\columnwidth]{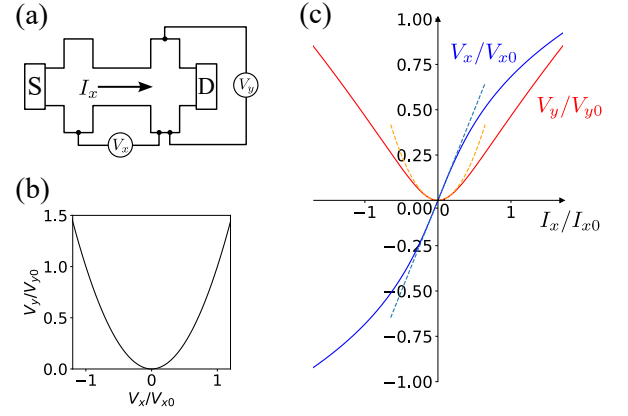}
    \caption{The longitudinal and Hall voltage $V_x, V_y$ as a function of the applied electric current $I_x$. (a) The Hall bar measurement. A current $I_x$ and a voltage $V_x$ are applied along $x$ direction to induce the Hall voltage $V_y$. (b) The resulting Hall voltage is quadratic in $V_x$, as shown in Eq.~\eqref{eq:ey}. (c) In contrast to (b), due to the bulk electric field effect ($\sigma_{xyx}$), $I_x$-$V_x$ deviates from linear relation (blue dashed line), and hence the $I_x$-$V_y$ curve deviates from the quadratic relation (orange dashed curve), as determined by Eqs.~\eqref{eq:asymmetric_tensor}, \eqref{eq:ey} and \eqref{eq:jxex}. Here, $V_{x0}=L_xE_0$, $V_{y0} = L_y \HPower E_0^2$, and $I_{x0}=(L_y L_z/L_x)\sigma_{xx} V_{x0}$.}
    \label{fig:open_circuit}
\end{figure}

When an applied voltage $V_x = E_x L_x$ drives a longitudinal current $I_x=j_x L_yL_z$ through a nonlinear Hall conductor of dimension $L_x \times L_y \times L_z$,  a transverse open-circuit voltage $V_y = -E_y L_y$ is induced by charge accumulation on the sides of the conductor due to NLHE. In order to obey the open-circuit condition with zero net current in $y$ direction, the drift current $j_d=\sigma_{yy} E_y$ driven by $E_y$ must exactly oppose and cancel the nonlinear Hall current $j_H=\sigma_{yxx} E_x^2$ driven by $E_x$. Solving Eq.~\eqref{eq:jy} under the condition $j_y=0$ yields a transverse electric field:   
\begin{eqnarray}
E_y = -\HPower E_x^2, \; {\rm with } \; \HPower = \frac{\sigma_{yxx}}{\sigma_{yy}}.  
\label{eq:ey}
\end{eqnarray}
The quadratic dependence of the induced transverse voltage $V_y$ on the applied voltage $V_x$ is the direct manifestation of second-order nonlinear Hall effect. At $E_x = \HPower^{-1}$, we have $E_y=E_x$ so that the nonlinear Hall angle is  \SI{45}{\degree}.

Interestingly, the induced transverse electric field $E_y$ in turn modifies the conductivity along $x$ direction through the $\sigma_{xyx}$ term. In addition to the $\sigma_{xxxx}$ term, this feedback effect causes the deviation of the longitudinal current-voltage relation from Ohm's law.  
Substituting Eqs.~\eqref{eq:ey} into \eqref{eq:jx} yields the longitudinal current-voltage relation: 
\begin{eqnarray}
j_x &=& \sigma_{xx} E_x + \qty(\sigma_{xxxx}+\frac{-\sigma_{xyx}\sigma_{yxx}}{\sigma_{yy}}) E_x^3. \nonumber \\
 &\equiv& \sigma_{xx} E_x \qty(1+ \frac{E^2_x}{E^2_0}). \label{eq:jxex}
\end{eqnarray}
Note that NLHE featuring $\sigma_{xyx}=-\sigma_{yxx}$ always leads to a decrease of $E_x$ induced by the applied current $j_x$, i.e., effectively decreases the longitudinal resistivity, which is opposite to Joule heating effect in metals. 
We defined a characteristic field strength $E_0= (\sigma_{xxxx}/\sigma_{xx}- \sigma_{yxx}\sigma_{xyx}/\sigma_{xx}\sigma_{yy})^{-\frac{1}{2}}$,  at which the resistivity operationally defined as $E_x/j_x$ is half of the linear response resistivity. Fig.~\ref{fig:open_circuit} shows the longitudinal current and transverse voltage as a function of input voltage, as determined by Eq.~\eqref{eq:ey} and Eq.~\eqref{eq:jxex}.

The nonlinear current-voltage characteristic enables frequency conversion. When an AC electric field $E_x(t) = E_x \cos(\omega t)$ is applied to the conductor, a time-dependent transverse voltage is generated. For $\omega \ll 1/\tau$ where $\tau$ is the scattering time of the conductor, $E_y(t)$ is simply determined by the instantaneous $E_x(t)$ through the quadratic relation Eq.~\eqref{eq:ey}: 
\begin{eqnarray}
E_y(t) = -\HPower E^2_x(t) = - \frac{\HPower E_x^2}{2} (1 + \cos(2\omega t)), \label{eq:Ey_Ex_squared}
\end{eqnarray}
which consists of a $2\omega$ component and a DC component.  A noncentrosymmetric conductor that generates a transverse DC voltage from the applied AC current or electric field through NLHE can be called ``Hall rectifier''. The coefficient $\HPower$, which measures the ratio of DC output voltage to AC input voltage, is an intrinsic material property. We call $\HPower$ the ``Hall power'', in analogy to thermopower that measures the voltage generated by temperature gradient.

The Hall rectifier can be utilized to deliver electric power to external load. Consider a setup shown in Fig.~\ref{fig:schematics_rectenna}, where the Hall rectifier is connected to a load resistance $R_L$.
Input power $\Pin$ flows into the device through the applied current $I_x$ or electric field $E_x$ in $x$-direction, and some of the input energy is transferred to the external load via the current $I_y$ to do useful work. The output power delivered to the load depends on both the nonlinear current-voltage characteristic of the Hall rectifier and the load resistance.  %
To calculate the output power, we note that the voltage across the load $V_y = -L_y E_y$ is related to $R_L$ and $I_y$ as $V_y = R_L I_y$, or  
equivalently, %
$j_y = -\sigma_L E_y$  with $\sigma_L \equiv L_y/(L_xL_z R_L)$. 
In the case of DC input, one obtains an expression for $E_y$ by solving this equation together with Eq.~\eqref{eq:jy} as   
\begin{align}
    E_y &= -\frac{\sigma_{yxx}}{\sigma_{yy}+\sigma_L}E_x^2. \label{eq:Ey_withEx} 
\end{align}
Substituting Eq~\eqref{eq:Ey_withEx} into \eqref{eq:jx}, the current in $x$-direction is obtained:  
\begin{align}
    j_x &= \sigma_{xx} E_x + \qty(\sigma_{xxxx} -\frac{\sigma_{xyx}\sigma_{yxx}}{\sigma_{yy}+\sigma_L})E_x^3. \label{eq:jxEx_load}
\end{align}
The relation between $j_x$ and $E_x$ reduces to the open circuit case ~\eqref{eq:jxex} at $\sigma_L\to0$ $(R_L\to\infty)$.

When the applied electric field is an AC field $E_x(t) = E_x \cos(\omega t)$ with $\omega\tau\ll 1$,  the induced longitudinal current $j_x(t)$ and transverse voltage $E_y(t)$ are also time-dependent and given by Eq~\eqref{eq:Ey_withEx} and Eq~\eqref{eq:jxEx_load} with the instantaneous field $E_x(t)$. 
The transverse voltage $V_y(t) \propto \cos^2(\omega t)$ now contains both a DC component and an AC component at frequency $2\omega$. Thus,  the electric power transferred to the load consists of both the DC power and the AC power: $W_L = W_L^\dc + W_L^\ac$ with  
\begin{align}
    W_L^\dc &= V_y^\dc I_y^\dc = -\vol E_y^\dc j_y^\dc, \label{eq:W_Ldc} \\
    W_L^\ac &= \overline{V_y^\ac I_y^\ac} = -\vol \overline{E_y^\ac j_y^\ac}, \label{eq:W_Lac}\\
    \Pin &= \overline{V_x I_x} = \vol \overline{E_x j_x}, \label{eq:Pin}
\end{align}
where $\overline{O(t)}$ represents the time average of $O(t)$. $E_y^\dc$ and $j_y^\dc$ are the DC components of $E_y$ and $j_y$, and similar for $E_y^\ac$ and $j_y^\ac$. 
Then one can calculate $W_L^\dc$, $W_L^\ac$, and $\Pin$ as 
\begin{align}
    W_L^\dc &= \vol \frac{\sigma_L\sigma_{yxx}^2}{4(\sigma_{yy}+\sigma_L)^2}E_x^4, \label{eq:W_Ldc_Ex}\\
    W_L^\ac &= %
    \frac{1}{2}W_L^\dc, \label{eq:W_Lac_Ex}\\ 
    \Pin &= \vol\qty[\frac{1}{2}\sigma_{xx} E_x^2 +\frac{3}{8}\qty(\sigma_{xxxx}+\frac{-\sigma_{xyx}\sigma_{yxx}}{\sigma_{yy}+\sigma_L})E_x^4]. \label{eq:Pin_ac_Ex}
\end{align}
These expressions for input and output power, derived from general current-voltage relation Eqs.~\eqref{eq:jx} and \eqref{eq:jy} for polar conductors, are exact up to the fourth order in the applied electric field $E_x$. The inclusion of additional terms in the current voltage relation will only add terms of sixth or higher order in $E_x$. Our analysis is applicable when $E_x$ is not too large so that higher-order corrections to the input and output powers are small.

The bulk electric field effect $\sigma_{xyx}$ plays an essential role in the transfer of input energy through the Hall rectifier to the load. It represents the additional input energy from the applied electric field $E_x$ to do external work on $R_L$ through the Hall voltage $-E_y L_y$.  
This can be seen when one considers changing the load resistance $R_L$ from $0$ (short-circuit) to a small finite value, while holding the applied electric field $E_x$ fixed. When $R_L=0$, the output power $I_y^2 R_L$ is zero despite the presence of a short-circuit transverse current $j_y = \sigma_{yxx} E^2_x$. Once $R_L$ becomes nonzero, an electric field $\delta E_y$ develops in the conductor, resulting in a  voltage $- \delta E_y L_y$ across the load which does useful work 
\begin{eqnarray}
\delta W_L / \vol = - \delta E_y j_y = -\sigma_{yxx} E^2_x \delta E_y.    
\end{eqnarray} 
At the same time, due to the bulk electric field effect $\sigma_{xyx}$, $\delta E_y$ results in an increase in the input current $\delta j_x =  \sigma_{xyx} E_x \delta E_y$, therefore increasing the input power by 
\begin{eqnarray}
\delta \Pin / \vol =  \delta j_x E_x = \sigma_{xyx} E^2_x \delta E_y.
\end{eqnarray}
Remarkably, thanks to the relation $\sigma_{xyx}=-\sigma_{yxx}$ in NLHE,  we find $\delta W_L = \delta \Pin$, i.e., the output power produced by small $\delta E_y$ is exactly compensated by the increase of input power. Similar consideration shows $\delta W_L= \delta \Pin$ when $R_L$ is changed from infinite (open-circuit) to a large finite value while the applied electric current $I_x$ is kept constant.

In general, a part of the input power $\Pin$, is transferred to the load as $W_L$, while the remaining part is dissipated as Joule heat $Q=\Pin - W_L$ in the Hall rectifier. We define the power transfer efficiency as $\eff_t\equiv W_L/\Pin$. Since the total power transferred to the load is a sum of DC and AC powers with $W_L^\dc=2W_L^\ac$, 
the efficiency of AC-DC power conversion or the rectification efficiency is given by $\eff_c\equiv W_L^\dc/\Pin=\frac{2}{3} \eff_t$.

It is instructive to first consider the hypothetical limit of a nonlinear Hall conductor without {\it any internal dissipation}, namely, $\sigma_{xx},\sigma_{yy}$ and $\sigma_{xxxx}$ are all zero, while $\sigma_{yxx}=-\sigma_{xyx}$ is finite. In this case, one can readily see that $\Pin=W_L^\dc+W_L^\ac$ holds for any load resistance $R_L$ and input field $E_x$, i.e., the power transfer efficiency $\eff_{t}$ is \SI{100}{\percent}. This is because the Hall current $\vec{j}^H$ is transverse to the electric field and does not induce any Joule heating. 

In reality, dissipation is inevitably present in the nonlinear Hall conductor. The power transfer efficiency $\eff_{t}$ is in general less than unity and depends crucially on the load resistance $R_L$ and the input power $\Pin$. 
The essential role of the load resistance and the input power on the energy transfer was not considered in a recent study on quantum rectification in noncentrosymmetric systems~\cite{shi_berry_2022}.

From Eqs.~\eqref{eq:W_Ldc_Ex}, \eqref{eq:Pin_ac_Ex}, one can readily see that $\eff_t$ increases monotonically with $E_x$, and at large $E_x$ compared to a certain value $E_1$, reaches 
\begin{align}
    \eff_{t} (E_x\gg E_1) 
    &= \frac{r_L}{(1+r_L)^2} \qty(\alpha + \frac{r_L}{1+r_L})^{-1},
    \label{eq:eff}
\end{align}
with $E_1$ given by 
\begin{align}
    E_1 
    &= \frac{2}{\sqrt{3}}E_0 \sqrt{\frac{1+\alpha}{\alpha + r_L/(1+r_L)}}
\end{align}
where $r_L=R_L/R_{yy}$ is the ratio of load resistance to the internal resistance $R_{yy}$.   $\alpha = \sigma_{xxxx}\sigma_{yy}/\sigma^2_{yxx} \geq 0$ is a dimensionless parameter that characterizes the strength of dissipative responses (both linear and higher-order) relative to NLHE, which is inherently dissipationless (${\bm j}^H \cdot {\bm E}=0$). A small $\alpha$ indicates a wide range of electric fields over which the current-voltage relation of the noncentrosymmetric crystal is dominated by NLHE. 
As an intrinsic material property, $\alpha$ can be experimentally determined by extracting $E_0$ and $\HPower$  from the current-voltage relation in the open circuit ($R_L=\infty$) and using the relation $\alpha=(\sigma_{xx}/\sigma_{yy}) (E_0 \HPower)^{-2}-1$.

When $\sigma_{xxxx}$ is small, the characteristic input power to obtain high efficiency is determined by intrinsic material parameters, given by 
\begin{align}
    P_0 &= \sigma_{xx}E_0^2 = \frac{\sigma_{yy}\sigma_{xx}^2}{\sigma_{yxx}^2}.
\end{align}
In order to achieve efficient rectification for energy harvesting applications, it is desirable to find nonlinear Hall materials with small $P_0$, so that dissipationless NLHE can dominate over dissipative responses at small input power.

By further maximizing $\eff_t$ with respect to $r_L$, we find the maximum power conversion efficiency 
\begin{align}
    \max\theta_t 
    &= \frac{1}{\qty(\sqrt{\alpha+1}+\sqrt{\alpha})^2},
\end{align}
which is attained at an optimum external resistance:
\begin{align}
    r^*_{L} =\sqrt{\frac{\alpha}{1+\alpha}}.
\end{align}
It should be noted that one can optimize $r_L$ by varying the length-to-width ratio of the nonlinear Hall conductor. For example, by increasing $L_y$ and decreasing $L_x$, one can increase $R_{yy}$ without changing the volume. Then, for the same input power $\Pin$ and external resistance $R_L$, $r_L$ is reduced so that the power conversion efficiency is increased for large $R_L$. 

\begin{figure}
    \centering
    \includegraphics[width=\columnwidth]{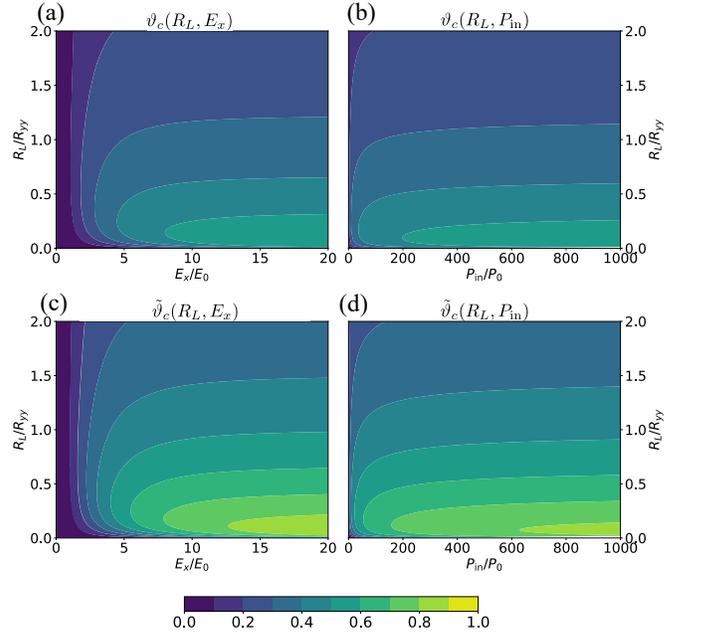}
    \caption{The efficiency for NLHE based rectification when $\alpha=0$. (a,b) $\eff_c$ for NLHE over $E_x$-$R_L$ plane (a) and $\Pin$-$R_L$ plane (b). (c,d) $\tilde{\eff}_c$ for NLHE when the large capacitance $C\gg (2\omega R_{yy})^{-1}$ is connected in parallel to the load resistance over $E_x$-$R_L$ plane (c) and $\Pin$-$R_L$ plane (d). Here, $P_0\equiv \sigma_{xx}E_0^2$. Because the capacitor acts as the low-pass filter, the energy loss due to AC current flowing into the load is suppressed and the efficiency can reach of order $1$. In both cases, high rectification efficiency is achieved when $R_L$ is small compared to $R_{yy}$ and $\Pin\gg P_0$ or $E_x\gg E_0$. }
    \label{fig:eff_Pin_RL_with_wo_capacitor}
\end{figure}

The maximum power transfer efficiency $\eff_{t}$ is a monotonically decreasing function of $\alpha$. As $\alpha\to0$ or $\sigma_{xxxx}\rightarrow 0$, the maximum  $\eff_{t}$ approaches $100\%$ at small $r_L$, hence the rectification efficiency $\eff_c$ approaches $2/3$. %
More realistically, for $r_L=1$ and assuming $\alpha=0$, we find $\eff_t=\SI{13.6}{\percent}$ at $E_x=E_0$, $\SI{45.2}{\percent}$ at $E_x=5E_0$, and $\SI{48.7}{\percent}$ at $E_x=10E_0$. 
If $\alpha=1$, it will be $\eff_t=\SI{6.0}{\percent}$ at $E_x=E_0$, $\SI{15.6}{\percent}$ at $E_x=5E_0$, and $\SI{16.4}{\percent}$ at $E_x=10E_0$.
Fig.~\ref{fig:eff_Pin_RL_with_wo_capacitor} shows the dependence of the efficiency $\eff_\ac$ on the applied electric field $E_x$ or input power $\Pin$ and on the external resistance $R_L$ for the case $\alpha=0$. 

In general, high efficiency is achieved when $\alpha$ is small and $E_x\gg E_1$, i.e., when NLHE ($\sigma_{yxx}, \sigma_{xyx}$) dominates over dissipative processes ($\sigma_{xx},\sigma_{yy},\sigma_{xxxx}$). Experimentally, this corresponds to large Hall angle. Indeed, when $\sigma_{xxxx}$ is negligible and $\sigma_{yxx}=-\sigma_{xyx}$, $\eff_t$ for $E_x\ll E_1$ can be rewritten with the Hall angle as 
\begin{align}
    \eff_t &= \frac{1}{1+r_L} \qty(\frac{1+r_L}{r_L} \frac{4(\sigma_{xx}/\sigma_{yy} )}{3\tan^2\Theta_H} + 1)^{-1}. \label{eq:eff_t2}
\end{align}
where $\Theta_H$ is the Hall angle, i.e., $\tan\Theta_H = E_y/E_x$.  Eq.~\eqref{eq:eff_t2} also shows that the large resistivity anisotropy $\sigma_{yy}/\sigma_{xx}\gg 1$ is helpful to achieve high efficiency.

Note that $1/3$ of the energy transferred to the load is in the form of AC power. This sets the upper limit of $2/3$ for AC-DC power conversion efficiency, when the Hall rectifier is directly connected to an external resistance. %
We can further increase the efficiency by suppressing the high-frequency $2\omega$  
components of the output current or voltage with a low-pass filter. This can be realized by connecting a capacitor to the load resistance in parallel. If the capacitance $C$ satisfies $(2\omega C)^{-1}\ll R_L, R_{yy}$, the capacitance approximately behaves as an open circuit for DC component but a short circuit for the $2\omega$ AC current. Hence, it is expected that most of the $2\omega$ component of $j_y$ does not flow into the load and does not consume energy, resulting in high AC-DC power conversion efficiency. Indeed, under this condition, 
we find  the rectification efficiency is equal to the power transfer efficiency, given by %
\begin{align}
    \tilde{\eff}_{c} &\simeq  \frac{\frac{r_L}{2(1+r_L)^2}E_x^2}{E_0^2 +\left( \frac{r_L}{2(1+r_L)}+ \frac{3}{4}\alpha \right) E_x^2 } \quad \textrm{for } 2\omega C \gg R_L^{-1}, R_{yy}^{-1}.
\end{align}
For a given $r_L$, it reaches maximum at large $E_x$ 
\begin{align}
    \max \tilde{\eff}_{c} = \frac{2r_L}{(1+r_L)(2r_L + 3\alpha(1+r_L))}. %
\end{align}
Fig.~\ref{fig:eff_Pin_RL_with_wo_capacitor}(c,d) shows the rectification efficiency when a large capacitance %
is connected to the Hall rectifier in parallel with the load resistance for $\alpha=0$. %
Now, the efficiency is much improved: it exceeds $2/3$  in broad region and reaches nearly $100\%$ at small $r_L$.

Instead of utilizing the low-pass filter, one can also extract work from the $2\omega$ components of $j_y$ by using another rectification device. If one utilizes NLHE again to rectify the $2\omega$ component, the maximum rectification efficiency will be $8/9\simeq \SI{89}{\percent}$.

To achieve efficient energy harvesting, impedance matching between the source for AC electricity and the rectifier is important~\cite{wagih_millimeter-wave_2020, brown_advances_2007}.
On this point, the Hall rectifier also has an advantage compared to the conventional diodes: one can choose its length and width $L_x$ and $L_y$ properly to match $R_x$ with the resistance of the antenna and at the same time optimize $R_y$ relative to the external resistance for high conversion efficiency. %

With our theory, we can estimate the rectification efficiency for the nonlinear Hall materials reported so far. A rough estimation of the efficiency for several materials is presented in the Supplemental Materials. For example, the estimated maximum conversion efficiency is \SI{4.3}{\percent} for bulk WTe$_2$ at the input current used in Ref.~\cite{shvetsov_nonlinear_2019}, assuming $\sigma_{yxx}=-\sigma_{xyx}$ and the isotropic resistance $R_{xx}=R_{yy}$. We expect higher efficiency by increasing the input power. 
When an additional mechanism (such as skew scattering) for nonlinear transverse response is involved, $\sigma_{yxx}\neq -\sigma_{xyx}$ in general and a more general expression for the power conversion efficiency can be found in Supplemental Materials.

Importantly, our Hall rectifier can work efficiently over a wide range of frequencies up to the scale of the scattering rate $1/\tau$ of the noncentrosymmetric crystal. Our analysis assumes that the linear and nonlinear conductivities do not depend on the frequency. This assumption is justified when the frequency $\omega$ is sufficiently small compared to the scattering rate $1/\tau$. Typically, $1/\tau$ of conductors at room temperature is on the order of \SI{10}{}-\SI{100}{T\hertz}, hence our theory of power conversion efficiency of Hall rectifiers is applicable over an ultrabroad frequency range from quasi-DC to the terahertz regime. In contrast, Schottky diodes have an upper cutoff frequency $\sim 1$THz set by $RC$ time constant of the semiconductor junction, while solar cells have a lower cutoff frequency $\sim 20$THz set by interband absorption threshold. %
In addition, the RF-DC conversion using the conventional Schottky diodes has limited efficiency for low input power $\sim\SI{}{\micro\watt}$~\cite{hemour_towards_2014}. In contrast, our Hall rectifier can work efficiently even for low input power.
Therefore, quantum rectification based on nonlinear Hall conductors may be a promising technology to harness RF, millimeter and THz waves for wireless charging and energy harvesting.

\acknowledgements
It is our pleasure to thank Zhiqiang Mao, Lujin Min and Suyang Xu for stimulating discussions and productive collaborations on related projects. This work is supported by the U.S. Army Research Laboratory and the U.S. Army Research Office through the Institute for Soldier Nanotechnologies, under Collaborative Agreement Number W911NF-18-2-0048. Y.O. is supported by JSPS KAKENHI Grant No.~JP22J22111. Y.O. thanks the support from Funai Overseas Scholarship. L.F. is partly supported by the David and Lucile Packard Foundation.

\bibliography{references, preprint}

\end{document}